\documentclass[a4paper,fleqn,twoside]{article}

\usepackage{espcrc2mm}
\usepackage{graphicx}
\usepackage{amsmath}
\usepackage{amssymb}

\title{Asymptotic normalization coefficient method for two-proton radiative
capture}

\author{%
L.V.~Grigorenko\address[dub]{Flerov Laboratory of Nuclear Reactions, JINR,
141980 Dubna, Russia}$^,$%
\address[mifi]{National Research Nuclear University ``MEPhI'', Kashirskoye
shosse 31, 115409 Moscow, Russia}$^,$%
\address[kur]{National Research Centre ``Kurchatov Institute'', Kurchatov sq.\
1, 123182 Moscow, Russia}$^,$%
\thanks{Corresponding author. E-mail address: lgrigorenko@yandex.ru (L.V.
Grigorenko)},
Yu.L.~Parfenova\addressmark[dub],
N.B.~Shulgina\addressmark[kur]$^,$%
\address[bog]{Bogoliubov Laboratory of Theoretical Physics, JINR, 141980 Dubna,
Russia},
M.V.~Zhukov\address[chal]{Department of Physics, Chalmers University of
Technology, S-41296 G\"{o}teborg, Sweden}
}

\begin{document}

\begin{abstract}
The method of asymptotic normalization coefficients is a standard approach for
studies of two-body non-resonant radiative capture processes in nuclear
astrophysics. This method suggests a fully analytical description of the
radiative capture cross section in the low-energy region of the astrophysical
interest. We demonstrate how this method can be generalized to the case of
three-body $2p$ radiative captures. It was found that an essential feature of
this process is the highly correlated nature of the capture. This reflects the
complexity of three-body Coulomb continuum problem. Radiative capture
$^{15}$O+$p$+$p \rightarrow ^{\,17}$Ne+$\gamma$ is considered as an
illustration.

\vspace{1.5mm}


\vspace{1.5mm}

\noindent \textit{Date:} \today.

\end{abstract}

\maketitle


\section{Introduction}
\label{sec:intro}


In the asymptotic normalization coefficient (ANC) approach the nuclear wave
function (WF) is characterized only by the behavior of its asymptotics. This
asymptotics is defined in terms of the modified Bessel function of the second
kind $K$ in neutral case
\[
\psi_{\text{gs}}(r \rightarrow \infty) = \mathcal{C}_{2} \,
\sqrt{2qr/\pi}\,  K_{l+1/2}(qr) \sim \mathcal{C}_{2} \exp[-qr] \,,
\]
or in terms of the Whittaker function $W$ in Coulombic case
\[
\psi_{\text{gs}}(r \rightarrow \infty) = \mathcal{C}_{2} \,
W_{-\eta,l+1/2}(2qr) \sim \mathcal{C}_{2} \, (2kr)^{-\eta} \exp[-qr] \,,
\]
where $\eta=Z_1Z_2 e^2 M/q$ is the Sommerfeld parameter. Thus the asymptotics
and, hence, the related observables are defined just by two parameters: the
g.s.\ binding energy $E_{\text{b}}=q^2/(2M)$ and the 2-body ANC value
$\mathcal{C}_{2}$.

Such an approximation is valid for highly peripheral processes. The nonresonant
radiative capture reactions at astrophysical energies are the main subject of
interest here
\cite{Xu:1994,Timofeyuk:2003a,Pang:2007,Okolowicz:2012,Mukhamedzhanov:2019a}. An
asymptotic normalization coefficient characterizes the virtual decay of a
nucleus into clusters and, therefore, it is equivalent to coupling constant in 
particle physics \cite{Blokhintsev:1977}. For that reason the ANC formalism
naturally provides a framework for deriving the low-energy astrophysical
information from peripheral reactions, such as direct transfer reactions, at
intermediate energies (the so-called ``Trojan horse'' method
\cite{Mukhamedzhanov:2006,Mukhamedzhanov:2017,Mukhamedzhanov:2019}). From the
short list of references above, it can be seen that the ANC study is quite
active and has a number of controversial unresolved issues.

For the network nucleosynthesis calculations in a thermalized stellar
environment it is necessary to determine the astrophysical radiative capture
rates $\left \langle \sigma _{\text{part}, \gamma } v \right \rangle$. The
two-body \emph{resonant} radiative captures
\begin{equation}
\langle \sigma_{\text{part},\gamma} v \rangle (T) \propto \frac{1}{T^{3n/2}}\,
\exp \left[ -\frac{E_r}{kT} \right]\, \frac{ \Gamma_{\gamma}
\Gamma_{\text{part}} } {\Gamma_{\mathrm{tot}}}\,,
\label{eq:res-rate-s}
\end{equation}
can be related to experimentally observable quantities
\cite{Fowler:1967,Angulo:1999,Grigorenko:2005a}: resonance position $E_r$, gamma
$\Gamma_{\gamma}$ and $\Gamma_{\text{part}}$ particle widths ($n=1$ for two-body
and $n=2$ for three-body captures).

The situation is much more complicated for \emph{nonresonant} radiative capture
rates. The direct measurements of the low-energy capture cross sections could be
extremely difficult for two-body processes. However, for the three-body capture 
rates the direct measurements of the corresponding capture cross sections are 
not possible at all. Therefore, experimental approaches to three-body processes 
include studies of the photo and Coulomb dissociation, which are reciprocal 
processes for radiative captures. However, the
``extrapolation'' of three-body cross sections from experimentally accessible
energies to the low energies, important for astrophysics, may require tedious
theoretical calculations. This is because relatively simple ``standard''
quasiclassical sequential formalism \cite{Fowler:1967,Angulo:1999} may not work
in essentially quantum mechanical cases
\cite{Grigorenko:2005a,Grigorenko:2006,Parfenova:2018,Grigorenko:2020b}.

The $2n$ and $2p$  astrophysical captures are becoming important at
extreme conditions in which density and temperature are so high  that triple 
collisions are possible. However, the temperature should not be too high to 
avoid the inverse  photodisintegration process. For the $2n$ captures the 
following possible astrophysical sites are investigated: (i) the neutrino-heated 
hot bubble between the nascent neutron star and the overlying stellar mantle of 
a type-II supernova, (ii) the shock ejection of neutronized material via 
supernovae, (iii) the merging neutron stars. The $2p$ captures may be important 
for explosive hydrogen burning in novae and X-ray bursts.

The $2n$ and $2p$ nonresonant radiative capture rates have been investigated in
a series of papers Ref.\ \cite{Grigorenko:2006,Parfenova:2018,Grigorenko:2020b}
by the examples of the $^{4}$He+$n$+$n \rightarrow ^{\,6}$He+$\gamma$ and
$^{15}$O+$p$+$p \rightarrow ^{\,17}$Ne+$\gamma$ transitions. These works also
required the development of exactly solvable approximations to understand
underlying physics of the process and achieve the accuracy needed for
astrophysical calculations
\cite{Grigorenko:2007,Grigorenko:2007a,Grigorenko:2018}. Some of the universal
physical aspects observed in the papers mentioned above have motivated the
search for simple analytic models. The following qualitative aspects of the
low-energy E1 strength function (SF) behavior were emphasized in
\cite{Grigorenko:2006,Parfenova:2018,Grigorenko:2020b} for $2n$ and $2p$ 
captures: (i) sensitivity to the g.s.\ binding energy $E_{\text{b}}$; (ii) 
sensitivity to the asymptotic weights
of configurations determining the transition; (iii) importance of one of
near-threshold resonances in the two-body subsystems (virtual state in $n$-$n$
channel in the neutral case and lowest resonance in the core-$p$ channel in the
Coulombic case), which effect on SF is found to be crucial even at
asymptotically low three-body energies. Points (i) and (ii) are the obvious
motivation for ANC-like developments; point (iii) represents important and
problematic difference from the two-body case.

This work to some extent summarizes this line of research suggesting analytical
framework for two-nucleon astrophysical capture processes. We demonstrate that
it is possible to generalize the two-body ANC2 method to the ANC3 method in the
situation of three-body radiative captures. While for the $2n$ capture the
practical applicability of ANC3 method remains questionable, for the $2p$
captures it is established beyond any doubt. In this work we provide compact
fully analytical framework for the processes, which previously could be
considered only in bulky numerical three-body calculations.


\section{ANC3 in the hyperspherical harmonics (HH) approximation}
\label{sec:anc3-hh}

\begin{figure}
\begin{center}
\includegraphics[width=0.49\textwidth]{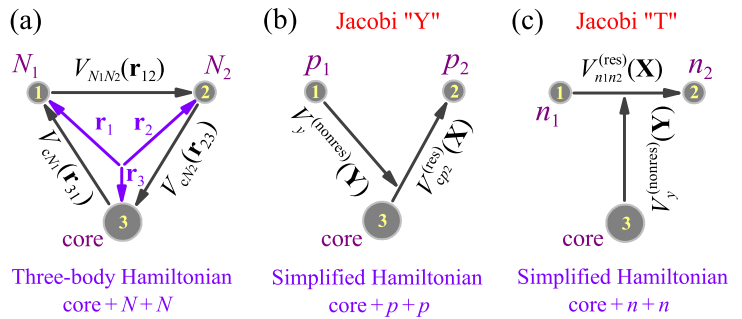}
\end{center}
\caption{Coordinate systems and potential sets for ``hyperspherical harmonics''
HH and ``simplified Hamiltonian'' SH approaches to ANC3. (a) The complete 3-body
Hamiltonian is applied both to core+$p$+$p$ and core+$n$+$n$ systems. (b) For 
core+$p$+$p$ system the dynamical domination of lowest
resonance in the core-$p$ subsystem motivates the use of simplified Hamiltonian
in the ``Y'' Jacobi system. (c) For core+$n$+$n$ system the dynamical domination
of the $n$-$n$ final state interaction motivates the use of simplified
Hamiltonian in the ``T'' Jacobi system.}
\label{fig:three-schemes}
\end{figure}

The HH formalism for calculations of the E1 SF is provided in details in Ref.\
\cite{Grigorenko:2020} and here we just give a sketch. Assume that the bound and
continuum wave functions (WF) can be described in a three-cluster core+$N$+$N$
approach by solving the three-body Schr\"odinger
equation
\begin{eqnarray}
(\hat{H}_3 - E_T)\Psi^J_{\text{gs}} = 0 \,, \nonumber \\
\hat{H}_3 = \hat{T}_3 + V_{N_1N_2}(\mathbf{r}_{12}) + V_{cN_2}(\mathbf{r}_{23}) 
+
V_{cN_1}(\mathbf{r}_{31}) + V_3(\rho) \,,
\label{eq:schred-bs}
\end{eqnarray}
where $E_T$ is the energy relative to the three-cluster breakup threshold. See
Fig.\ \ref{fig:three-schemes} for definition of coordinates used in this work. 
The pairwise interactions $V_{ij}$ are motivated by spectra of the subsystems, 
while $V_3$ is phenomenological three-body potential used for fine-tuning of the 
three-body resonance energies.
In the hyperspherical harmonics method this equation is reduced to a set of
coupled differential equations
\begin{eqnarray}
\Psi^J_{\text{gs}}(\rho,\Omega_5) = \rho^{-5/2} \sum \nolimits _{K \gamma} 
\chi_{K
\gamma}(\rho) \mathcal{J}_{JK \gamma} (\Omega_5) \,, \nonumber \\
\left [ \frac{d^2}{d \rho^2} - \frac{\mathcal{L}(\mathcal{L}+1)}{\rho^2} +
2M(E_T-V_{K \gamma,K \gamma}(\rho)) \right ] \chi_{JK \gamma}(\rho) \nonumber \\
= \, \sum \nolimits _{K' \gamma' \neq K \gamma} 2M \, V_{K' \gamma',K 
\gamma}(\rho) \, \chi_{JK' \gamma'}(\rho) \,,
\label{eq:hh-se-decomp} \\
\rho ^{2} = (A_1 A_2 r_{12}^{2}+ A_2 A_3 r_{23}^{2} + A_3 A_1 r_{31}^{2}) / (A_1
+ A_2+ A_3)  \,, \nonumber
\end{eqnarray}
depending on the collective coordinate --- hyperradius $\rho$. The ``scaling''
mass $M$ is taken as an average nucleon mass in the system and $\mathcal{J}_{JK
\gamma} (\Omega_5)$ is the hyperspherical harmonic with the definite total spin 
$J$. The three-body potentials are defined as
\[
V_{K' \gamma',K \gamma}(\rho) = \langle \mathcal{J}_{JK' \gamma'} | \textstyle
\sum \nolimits _{i > j} V_{ij}(\mathbf{r}_{ij}) | \mathcal{J}_{JK \gamma}
\rangle \,.
\]
The effective orbital momentum $\mathcal{L}=K+3/2$ is nonzero even for the
lowest excitation $K=0$.

The continuum three-body problem is solved using the same Eq.\
(\ref{eq:hh-se-decomp}) set but for continuum WF
$\chi_{JK_{f}\gamma_{f},K_{f}^{\prime}\gamma
_{f}^{\prime}}(\varkappa \rho)$ (square matrix of solutions)
diagonalizing S-matrix. Hypermomentum $\varkappa$ is defined as
$\varkappa=\sqrt{2ME_T} $. In the no-Coulomb case the WF is constructed by
diagonalizing the $3 \rightarrow 3$ elastic scattering S-matrix on asymptotics
\begin{eqnarray}
\chi_{J K \gamma,K' \gamma'}(\varkappa \rho) =
\exp(i \delta_{K \gamma,K' \gamma'})
\sin(\varkappa \rho - (K+2) \pi/2
\nonumber \\
+ \delta_{K \gamma,K'\gamma'}) \,,\qquad S_{K \gamma,K' \gamma'}  =  \exp(2i
\delta_{K \gamma,K' \gamma'}) \,,
\nonumber
\end{eqnarray}
in analogy with the two-body case. This WF contains plane three-body wave and 
outgoing waves. The formulation of the boundary conditions becomes problematic 
in the Coulomb case and methods with only outgoing waves (including the SH model 
introduced later in Section \ref{sec:sh}) is a preferable choice. The details of 
the method and its applications are well explained in the literature 
\cite{Zhukov:1993,Grigorenko:2001,Grigorenko:2005a,Grigorenko:2009c,%
Pfutzner:2012,Grigorenko:2020} and we will not dwell on that too much.

The form of hyperspherical equations (\ref{eq:hh-se-decomp}) immediately
provides the vision for the low-energy behavior of observables in E1 continuum
since the only $K=1$ component with the lowest centrifugal barrier is important
in the $E_T \rightarrow 0$ limit.

The E1 transitions between three-body cluster core+$N$+$N$ states are induced
by the following operator
\[
\mathcal{O}_{\text{E1},m}=e \, \sum \nolimits _{i=1,3}
Z_{i}\, r_{i} \,Y_{1 m}(\hat{r}_{i})=\sqrt{3/(4 \pi)}  \, D_{m} \,,
\]
where $\mathbf{D} = \textstyle \sum_{i=1,3} e Z_{i} \mathbf{r}_{i}$ is the
dipole operator, and
\begin{eqnarray}
\mathcal{O}_{\text{E1},m} =
Z_{\text{eff}}\, \rho \, \cos(\theta_{\rho}) \, Y_{1m}(\hat{y}) \,,
\label{eq:trans-oper}  \\
Z_{\text{eff}}^{2}=
\left \{
\begin{array}{l}
Z^2_{3}  \\
(Z_{3}-A_{3})^2
\end{array}
\right \}
 \frac{e^2 \,(A_{1}+A_{2})}{A_{3}(A_{1}+A_{2}
+A_{3})} \, .
\label{eq:z-eff-hh}
\end{eqnarray}
The upper value in curly braces is for core+$n$+$n$ and the lower one is for
core+$p$+$p$ three-body systems, taking into account the c.m.\ relation
$\mathbf{r}_{1}+\mathbf{r}_{2}= -A_{3}\mathbf{r}_{3}$ for the three-body system.

For historical reasons the astrophysical E1 nonresonant radiative \emph{capture} 
rate is expressed via the SF of the reciprocal E1 \emph{dissociation}, see Eq.\ 
(\ref{eq:nonres-rate}). The E1 dissociation SF in the HH approach is
\begin{equation}
\frac{dB_{E1}}{dE_T} =
\sum \nolimits_{J_f} G_{fi} \, \sqrt{\frac{M}{2E_T}}
\sum \nolimits_{K_f \gamma_f} |M_{J_f K_f \gamma_f}|^2 \,.
\label{eq:dbe1-def}
\end{equation}

where $J_i$ is total spin of bound state, $J_f$ is total spin of continuum 
state, $G_{fi}=(2J_f+1)/(2J_i+1)$ is a statistical factor, and the E1 matrix 
element is
\begin{eqnarray}
M_{J_{f}K_{f}\gamma_{f}}=Z_{\text{eff}}\,
{\sum \nolimits_{K_{f}^{\prime}\gamma_{f}^{\prime}}}
{\sum \nolimits_{K_{i}\gamma_{i}}}M_{\text{a}} M_{\text{hh}}
\int d \rho \, \sqrt{2/\pi} \nonumber \\
\times \, \chi_{J_{f} K_{f} \gamma_{f},K_{f}^{\prime}\gamma
_{f}^{\prime}}(\varkappa \rho) \, \rho \, \chi_{J_{i}K_{i}\gamma_{i}}(\rho)\,,
\nonumber \\
M_{\text{a}} = \left \langle J_{f} \gamma_{f}^{\prime}\left \Vert
Y_{1}(\hat{y})\right \Vert J_{i} \gamma_{i} \right \rangle \,,\quad
M_{\text{hh}}=\left \langle K_{f}^{\prime}\gamma_{f}^{\prime} \left \vert
\cos(\theta_{\rho})\right \vert K_{i}\gamma_{i} \right \rangle  \,. \nonumber
\end{eqnarray}
For example, the reduced angular momentum matrix element $M_{\text{a}} =
1/\sqrt{4 \pi}$ for $J_i=0 \rightarrow J_f=1$ transition and the hyperangular 
matrix
element $M_{\text{hh}} = 1/\sqrt{2}$ for $K_i=0 \rightarrow K_f=1$ transition.


\subsection{No Coulomb case in HH approach}
\label{sec:anc3-hh-nc}


\begin{figure}
\begin{center}
\includegraphics[width=0.49\textwidth]{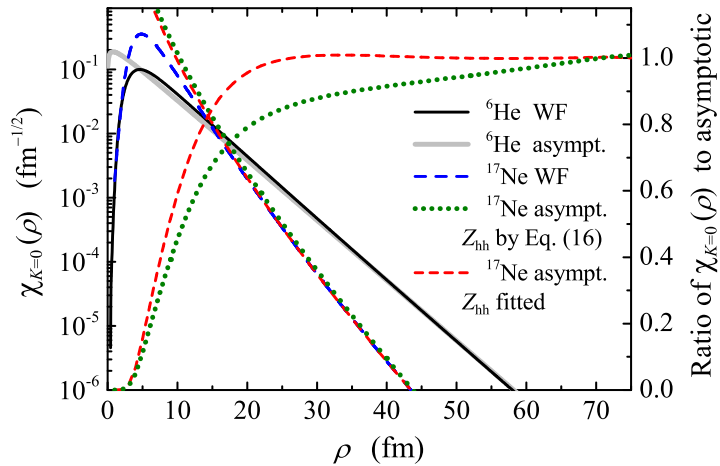}
\end{center}
\caption{Left axis: ground state WF components with $K=0$ for $^{6}$He (solid
curve) and $^{17}$Ne (dashed curve), matched to asymptotic Bessel (thick gray
curves) and Whittaker (short dashed and dotted curves) functions. For $^{6}$He
ANC3 value is $\mathcal{C}_{3}=0.3$ $\text{fm}^{-1/2}$. For $^{17}$Ne ANC3
values are $\mathcal{C}_{3}=13400$ $\text{fm}^{-1/2}$ with $Z_{\text{hh}}=27.5$
[see Eq.\ (\ref{eq:3b-coul-effc})] and $\mathcal{C}_{3}=5958$ $\text{fm}^{-1/2}$
with fitted $Z_{\text{hh}}=26.14$. The lines on the right axis show the ratio of
$^{17}$Ne WF to Whittaker functions with mentioned $Z_{\text{hh}}$ and
$\mathcal{C}_{3}$.}
\label{fig:wfs-vs-ass}
\end{figure}

For the three-body plane-wave case the solution matrix is diagonal and expressed
in terms of cylindrical Bessel functions
\[
\chi_{J_f K_{f}\gamma_{f},K_{f}^{\prime}\gamma
_{f}^{\prime}}(\varkappa \rho) = \sqrt{\frac{2}{\pi}} \, \delta_{ K_f,
K_f^{\prime}} \, \delta_{\gamma_{f}, \gamma_{f}^{\prime}} \, \sqrt{\frac{\pi
\varkappa \rho}{2}} \, J_{K_f+2}(\varkappa \rho) \,,
\]
with asymptotics for small $\varkappa \rho$
\begin{equation}
\sqrt{\varkappa \rho} \, J_{K_f+2}(\varkappa \rho) \sim
(\varkappa \rho)^{K_f+5/2}/[(K_f+2)! \, 2^{K_f+2}] \,.
\label{eq:bess-ass}
\end{equation}
This expression can be used to separate the leading term of the low-energy
dependence of the matrix element, labeled for simplicity only by the values of $
K $ for the initial and final states
\begin{eqnarray}
M_{K_f K_i}(E_T) & = & \sqrt{2/\pi} \, \varkappa^{K_f+5/2} Z_{\text{eff}}
M_{\text{a}} M_{\text{hh}}   I_{K_fK_i}(E_T)\, ,
\label{eq:m-sep} \\
I_{K_f K_i}(E_T) & = &   \frac{1}{\varkappa^{K_f+2}} \int d \rho \,
 \rho^{3/2} \, J_{K_f+2}(\varkappa \rho) \, \chi_{J_{i}K_{i}\gamma_{i}}(\rho)
 \,. \qquad
\label{eq:ikf}
\end{eqnarray}
where the overlap integral $I_{K_f K_i}$ tends to a constant at $E_T \rightarrow
0$ and weakly depends on energy in the range of interest.

Let us consider only the transition from the lowest bound state component
$K_i=0$ to the lowest E1 continuum component with $K_f=1$:
\begin{equation}
\frac{dB_{E1}}{dE_T} = \frac{1}{\pi} \, G_{fi} \,
Z^2_{\text{eff}}M^2_{\text{a}} M^2_{\text{hh}} \, (2M)^{4} \, E_T^{3} \,
I^2_{10}(E_T) \,.
\label{eq:dbde-nc}
\end{equation}
Now we replace the bound state WF $\chi$ in Eq.\ (\ref{eq:ikf}) by its
long-range asymptotics expressed in terms of the three-body ANC value
$\mathcal{C}_{3}$ and cylindrical Bessel functions $K$
\begin{equation}
\chi_{K=0}(\rho) \rightarrow \mathcal{C}_{3} \, \sqrt{2 \kappa \rho/\pi}
\, K_{2}(\kappa \rho)\,,
\end{equation}
where the g.s.\ hypermoment $\kappa = \sqrt{2ME_{\text{b}}}$ is defined via the
binding energy $E_{\text{b}}$. This approximation is valid in a broad range of
$\rho$ values, see Fig.\ \ref{fig:wfs-vs-ass}. The $^{6}$He WF is taken from
\cite{Grigorenko:2020,Grigorenko:2020b}. The overlap integral now has simple
analytical form
\begin{equation}
I_{10}(E_T) =  4\, \mathcal{C}_{3} \, / [(2ME_{\text{b}})^{11/4}
(1+E_T/E_{\text{b}})^2] \,.
\label{eq:i2pw-k0}
\end{equation}
It can be found that the ANC3 approximation of the overlap value
(\ref{eq:i2pw-k0}) deviates within very reasonable $\sim 7\%$ limits from the
directly calculated by Eq.\ (\ref{eq:ikf}) in a broad energy range ($E_T
\lesssim 1$ MeV).


\subsection{Discussion of $^{6}$He case}


The E1 SF and the astrophysical capture rate for the $\alpha$+$n$+$n \,
\rightarrow \,^{6}$He+$\gamma$ was recently studied in Refs.\
\cite{Grigorenko:2020b,Grigorenko:2020}. It can be found that Eq.\
(\ref{eq:dbde-nc}) is not sufficient in this case for two reasons:

\noindent (i) In the $p$-shell $^{6}$He nucleus not only the $K_i=0
\rightarrow K_f=1$ transition is important, but also $K_i=2 \rightarrow K_f=1$.
The asymptotics of the $K_i=2$ WF component falls off much faster than that of
the component $K_i=0$. However, the weight of the $K_i=2$ WF component
corresponding to $[p^2]$ configuration is much larger ($\sim 80 \%$), than the
weight of the $K_i=0$ WF component ($\sim 5 \%$), due to Pauli-suppressed
$[s^2]$ configuration. So, finally their contributions to the low-energy ME are
comparable.

\noindent (ii) It was shown in \cite{Grigorenko:2020b,Grigorenko:2020} that the
low-energy part of the E1 SF is highly sensitive to the final state $n$-$n$
interaction (an increase in SF when the $n$-$n$ interaction is taken into
account is a factor of 8). The paper \cite {Grigorenko:2020b} is devoted to the
study of this effect in the dynamic dineutron model. We do not currently see a
method to consider this effect analytically.

Applicability of the approximation (\ref{eq:dbde-nc}) to the other cases of $2n$
capture should be considered separately.


\subsection{Coulomb case in HH approach}


Let us consider the transition to the single $K_f=1$ continuum final state. The
low-energy behavior of continuum single channel WF in the
Coulomb case is provided by the regular at the origin Coulomb WF
\begin{equation}
\chi_{K_{f}}(\varkappa \rho) \rightarrow \sqrt{2/\pi} \,
F_{K_{f}+3/2}(\eta_{\text{hh}},\varkappa \rho)\,.
\label{eq:coul-ass-0}
\end{equation}
The suitable asymptotics of the Coulomb WFs are
\begin{eqnarray}
F_{l}(\eta,kr) & = & \frac{(2l+1)! C_{l}(\eta)}{(2 \eta)^{l+1}} \sqrt{2 \beta r}
\, I_{2l+1}(2 \sqrt{2 \beta r}) \,,
\label{eq:coul-ass} \\
G_l(\eta,kr) & = & \frac{2(2 \eta)^l}{(2l+1)!C_l(\eta)} \sqrt{2 \beta r}
\,K_{2l+1}(2\sqrt{2 \beta r}) \,,
\label{eq:coul-g-ass} \\
C_{l} (\eta) & = & 2^l \exp[-\pi \eta/2] \, \left \vert \Gamma[l + i \eta + 1]
\right \vert / \Gamma[2 (l + 1)]  \, , \qquad
\label{eq:coul-const} \\
C'_{l} (\eta) & = & \frac{\sqrt{\pi}(2 \eta)^{l+1/2} } {(2l+1)!} \, \exp[-\pi
\eta ]  \,,
\qquad \beta = \eta k  \, ,
\label{eq:coul-const-approx} \\
D_l(\eta,k) & = & \frac{(2l+1)! C'_{l}(\eta)}{(2 \eta)^{l+1}} \sqrt{2 \beta}
= \sqrt{\pi k} \, \exp[-\pi \eta] \,, \qquad
\label{eq:d-coef}
\end{eqnarray}
where $I$ and $K$ are modified Bessel functions. Approximation
(\ref{eq:coul-const-approx}) for the Coulomb coefficient (\ref{eq:coul-const})
works for $\eta \gg l$.

\begin{figure*}
\begin{center}
\includegraphics[width=1.00\textwidth]{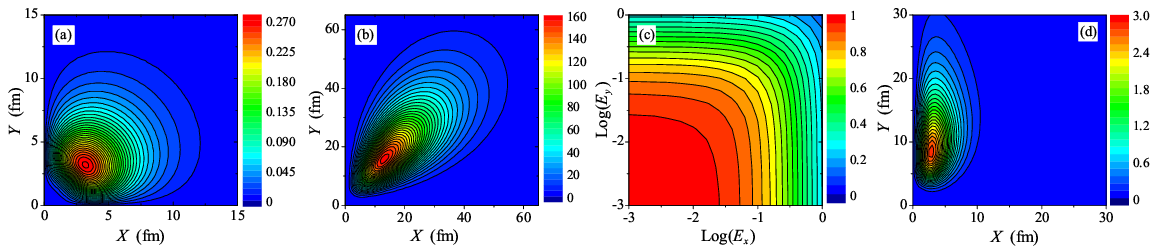}
\end{center}
\caption{(a) The WF $\psi_{\text{gs}}$ component with $K=0$, normalized to
unity. (b) Integrand of Eq.\ (\ref{eq:ikfcc-asym}). (c)
$I^{(cc)}_{l_xl_y}(\varepsilon_0,E_T)/\tilde{I}^{(cc)}_{l_xl_y}$ on logarithmic
axes. (d) Integrand of Eq.\ (\ref{eq:b-int}). }
\label{fig:wfs-xy}
\end{figure*}

In the ANC3 approximation the g.s.\ WF $\chi$ can be replaced by its
long-range asymptotics
\begin{equation}
\chi_{K_i}(\rho) \rightarrow \mathcal{C}_{3} \,  W_{-\eta_{\text{gs}},K_i+2}(2
\kappa \rho)  \,.
\label{eq:3b-gs-as}
\end{equation}
This asymptotics is valid when all three particles are well separated. We will 
find out later that at least the core-$p$ distances, which contribute E1 SF, are 
simultaneously large, see Fig.\ \ref{fig:wfs-xy} (b). The Sommerfeld parameters 
$\eta$ for continuum and bound states are
\begin{equation}
\eta_{\text{hh}} = \beta_{\text{hh}} /\varkappa \,,
\quad \beta_{\text{hh}} = Z_{\text{hh}}e^2  M \,,
\quad  \eta_{\text{gs}} = Z_{\text{hh}} \, e^2  M/\kappa \,.
\label{eq:sommer-z-hh-eff}
\end{equation}
The effective charges of isolated hyperspherical channels can be defined as
\begin{equation}
Z_{\text{hh}}^{(K l_x l_y)} = \rho  \left \langle K l_x l_y \left \vert
\frac{Z_1 Z_2}{r_{12}} + \frac{Z_2 Z_3}{r_{23}} + \frac{Z_3 Z_1}{r_{31}} \right
\vert K l_x l_y \right \rangle .
\label{eq:3b-coul-effc}
\end{equation}
For the $^{17}$Ne case the $K=0$ and $K=1$ effective charges are
\begin{equation}
Z_{\text{hh}}^{(000)} =  27.50 \, , \quad Z_{\text{hh}}^{(101)} =  27.41 \, .
\label{eq:zhh-k0-k1}
\end{equation}
Fig.\ \ref{fig:wfs-vs-ass} shows that the substitution Eq.\ (\ref{eq:3b-gs-as})
works well in a very broad range of radii (the $^{17}$Ne g.s.\ WF is from Ref.\
\cite{Grigorenko:2003}). The effective charge in Eq.\
(\ref{eq:zhh-k0-k1}) obtained for $K_i=0$ is very reasonable. However, slightly
different effective charge value $Z_{\text{hh}}^{(000)} =26.14$ is required for
an almost perfect match to the asymptotics. This is a clear indication of
coupled-channel dynamics in this case. It is actually a nontrivial fact that all
the complexity of this dynamics reduces to a simple renormalization of effective
charges.

\begin{figure}
\begin{center}
\includegraphics[width=0.45\textwidth]{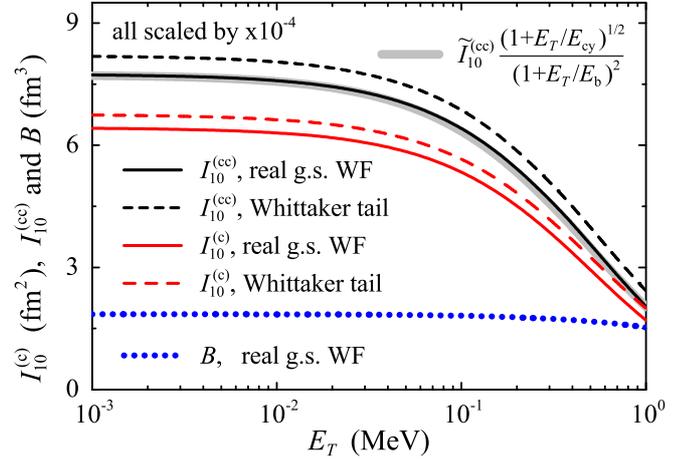}
\end{center}
\caption{Overlap integrals for HH $I^{(c)}_{10}(E_T)$ [Eq.\ (\ref{eq:ikfc})],
SH $I^{(cc)}_{10}(\varepsilon_0,E_T)$ [Eq.\ (\ref{eq:ikfcc})], and
$B(\varepsilon_0,E_T)$ [Eq.\ (\ref{eq:b-int})]. The solid gray curve shows
analytical approximation Eq.\ (\ref{eq:icc-ener-dep}) for $I^{(cc)}_{10}$.}
\label{fig:overlap-int}
\end{figure}

\begin{figure}
\begin{center}
\includegraphics[width=0.46\textwidth]{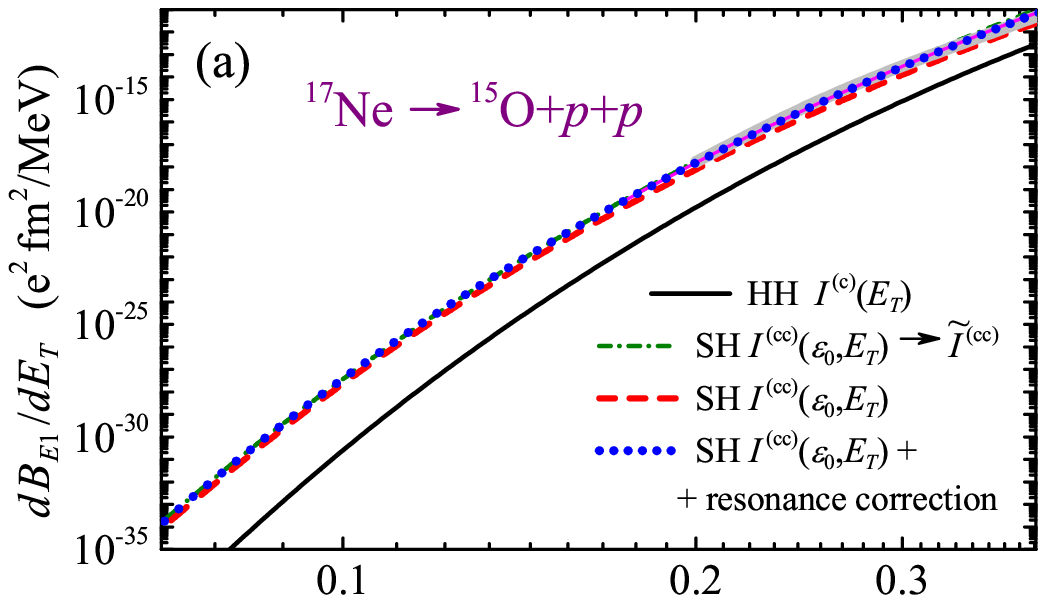} \\
\includegraphics[width=0.46\textwidth]{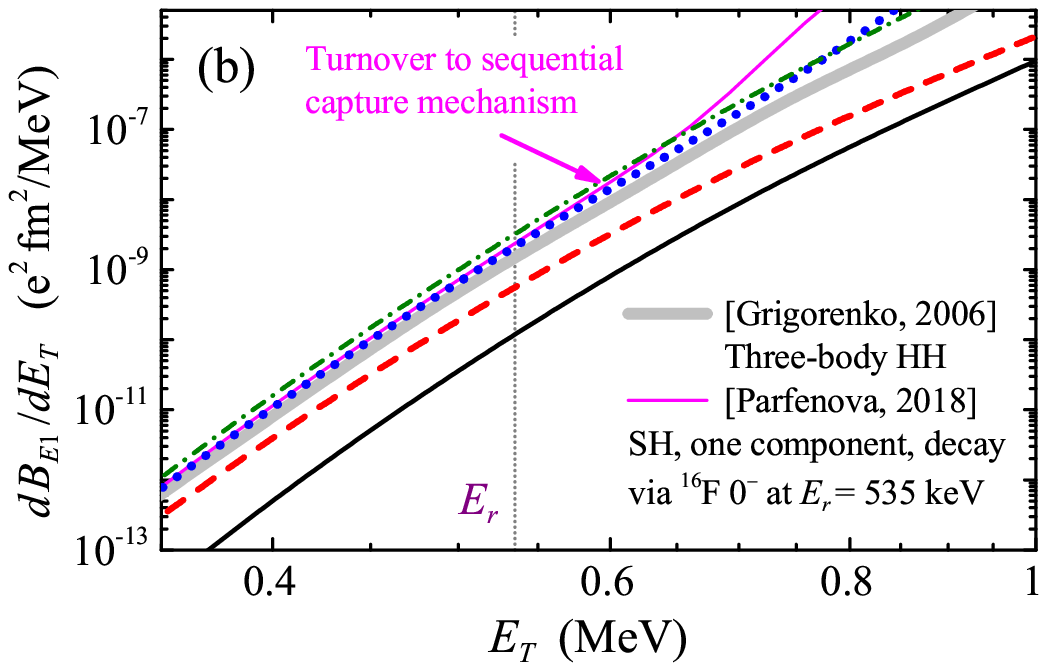}
\end{center}
\caption{The E1 strength functions for $^{17}$Ne $\rightarrow ^{15}$O+$p$+$p$
transition. Solid black curve corresponds to E1 SF obtained with ANC3 HH method
of Eq.\ (\ref{eq:dbde1-cou-hh}). Green dash-dotted curve corresponds to simple
energy-independent approximation Eq.\ (\ref{eq:icc-corr})
$I^{(cc)}(\varepsilon,E_T) \rightarrow \tilde{I}^{(cc)}$. Red dashed and blue
dotted curves corresponds to ANC3 SH method of Eq.\ (\ref{eq:dbde2-cou-sh})
without and with the resonance correction Eq.\ (\ref{eq:icc-corr}), 
respectively. Thick gray curve and thin magenta solid curves correspond to SFs 
from Refs.\ \cite{Grigorenko:2006} and \cite{Parfenova:2018}, respectively.}
\label{fig:e1-sf-17ne}
\end{figure}

Using Eqs.\ (\ref{eq:coul-ass}) and (\ref{eq:d-coef}) we can factorize the $E1$
matrix element as:
\begin{eqnarray}
M_{K_f K_i}  =  \sqrt{2/\pi} \, D_{K_f+3/2}(\eta_{\text{hh}},\varkappa) \,
M_{\text{a}}M_{\text{hh}} Z_{\text{eff}} \, I^{(c)}_{K_{f} K_{i} }(E_T)\,,
\nonumber \\
I^{(c)}_{K_f K_i}(E_T)  =  \int d \rho \,
\frac{F_{K_{f}+3/2}(\eta_{\text{hh}},\varkappa
\rho)}{D_{K_f+3/2}(\eta_{\text{hh}},\varkappa)} \, \rho^{3/2}
\, \chi_{J_{i}K_{i}\gamma_{i}}(\rho)\,, \qquad
\label{eq:ikfc}
\end{eqnarray}
where the overlap integral $I^{(c)}_{K_f K_i}$  weakly depends on the energy
and in the limit $E_T \rightarrow 0$ has a rather simple form
\begin{equation}
\tilde{I}^{(c)}_{K_f K_i} = \int d \rho \, I_{2l+1}(2 \sqrt{2 \beta_{\text{hh}}
\rho})\, \rho^{3/2} \, \chi_{J_{i}K_{i}\gamma_{i}}(\rho)\,.
\label{eq:ikfc-asymp}
\end{equation}
The overlaps (\ref{eq:ikfc}) for $K_i=0 \rightarrow K_f=1$ transition are shown
in Fig.\ \ref{fig:overlap-int}. It can be found that in the ANC3 approximation
the Eq.\ (\ref {eq:3b-gs-as}) is quite accurate: in this case the overlap
increases just less than $6 \%$ compared the calculation with the real g.s.\ WF.
It is also seen that the use of simple energy-independent overlap Eq.\
(\ref{eq:ikfc-asymp}) instead of (\ref{eq:ikfc}) gives almost perfect result
below 10 keV and is reasonable below 100 keV. For the $E1$ SF we get:
\begin{equation}
\frac{dB_{E1}}{dE_T} = G_{fi} \, Z^2_{\text{eff}} M^2_{\text{a}} M^2_{\text{hh}}
2M I^{(c)2}_{10}(E_T) \exp[-2 \pi \eta_{\text{hh}}]\, .
\label{eq:dbde1-cou-hh}
\end{equation}
The energy dependence of the derived expression at $E_T \rightarrow 0$ is pure
Coulomb exponent $\exp[-2 \pi \eta_{\text{hh}}]$. The SF calculation results are
shown in Fig.\ \ref{fig:e1-sf-17ne}. They strongly disagree with calculation
results from Refs.\ \cite{Grigorenko:2006} and \cite{Parfenova:2018}. The
modification of the ``effective continuum charge'' $Z_{\text{hh}}^{(101)}$ from
Eq.\ (\ref{eq:zhh-k0-k1}) does not save the situation since the energy
dependence of the SF in Eq.\ (\ref{eq:dbde1-cou-hh}) and that of the SF in
\cite{Grigorenko:2006,Parfenova:2018} are too different. We demonstrate in the
next section that the Eq.\ (\ref{eq:dbde1-cou-hh}) is actually incorrect.
However, the derivations of this section are still important for our further
discussion.

\section{ANC3 in the simplified Hamiltonian (SH) approximation}
\label{sec:sh}

The approximation is based on the usage of a simplified three-body Hamiltonian
for the E1 continuum instead of the real one
\begin{equation}
\hat{H}_3 \rightarrow \hat{H}'_3 = \hat{T}_3 + V_{cN_2}(\mathbf{X}) +
V_{y}(\mathbf{Y}) \,,
\label{eq:sh}
\end{equation}
where $\mathbf{X} \equiv \mathbf{r}_{32}$ is the Jacobi vector in the ``Y''
Jacobi system, while $\mathbf{Y}$ corresponds to the second Jacobi vector, see
Fig.\ \ref{fig:three-schemes}. Such a Hamiltonian is quite reliable since the
nuclear interaction with a proton in a non-natural parity state is weak.
The model was used for nonresonant astrophysical rate calculations in $^{17}$Ne
in Ref.\ \cite{Grigorenko:2006} and in $^{6}$He in Ref.\
\cite{Grigorenko:2020b}. A thorough check of the model is given in Ref.\
\cite{Grigorenko:2007}, and the detailed description of the formalism for
complicated angular momentum couplings in Ref.\ \cite{Parfenova:2018}.

To obtain the E1  \emph{dissociation} strength function in this approximation we 
solve the inhomogeneous Schr\"odinger equation
\[
(\hat{H}'_3-E_T )
\Psi_{M_im}^{J_{f}M_f(+)}=\mathcal{O}_{\text{E1},m}\Psi^{J_{i}M_i}_{\text{gs}}
\,,
\]
for WF $\Psi^{(+)}$ with pure outgoing wave boundary conditions.
The transition operator Eq.\ (\ref{eq:trans-oper}) dependent on $\mathbf{r}_3 $
can be rewritten in $X$ and $Y$ coordinates using relation:
\begin{equation}
\mathbf{r}_3 = \mathbf{X}\, A_2/(A_2+A_3)  - \mathbf{Y}\, A_1/(A_1+A_2+A_3)
 \, .
\label{eq:r3-xy}
\end{equation}
Since the factorized form of the Hamiltonian Eq.\ (\ref{eq:sh}) allows a
semi-analytical expression for the three-body Green's function, a rather simple
expression for the SF can be obtained
\begin{eqnarray}
\frac{dB_{E1}}{dE_T}  = G_{fi} \, \frac{4}{\pi^2} \, E_T \int_0^{1} d
\varepsilon \,
\frac{M_xM_y}{k_xk_y} \, |A(E_x,E_y)|^2 \,, \qquad
\label{eq:flux-sh}   \\
E_x =\varepsilon E_T \,, \quad E_y =(1-\varepsilon) E_T  \,, \quad k_{x,y}
=\sqrt{2M_{x,y}E_{x,y}} \,, \nonumber
\end{eqnarray}
where $\varepsilon$ is the energy distribution parameter. The amplitude $A$ is
defined as
\begin{equation}
A(E_x,E_y) = \int dX dY \, f_{l_x}(k_xX) \, f_{l_y}(k_yY) \, \Phi(X,Y) \,.
\label{eq:amp-ex}
\end{equation}
where the ``source function'' $\Phi$ is defined by the E1 operator acting on
$\Psi_{\text{gs}}$. The WFs $f_{l_x}$ and $f_{l_y}$ are eigenfunctions of
sub-Hamiltonians depending on $X$ and $Y$ Jacobi coordinates in S-matrix
representation with asymptotics
\begin{equation}
f_l(kr) = e^{i \delta_l}[ F_l(\eta, kr) \cos(\delta_l) + G_l(\eta, kr)
\sin(\delta_l)]\,.
\label{eq:con-wf}
\end{equation}

Eq.\ (\ref{eq:amp-ex}) is given in a simplified form, neglecting angular
momentum couplings, more details can be found in \cite{Parfenova:2018}. We skip
this part of the formalism in this work. The calculations of the E1 strength
function in the SH approximation without final state interactions in $X$ and $Y$
channels for the $2n$ capture are equivalent to calculations in the HH
approximation. So, we skip no-Coulomb case and proceed to the $2p$ capture.


\subsection{Coulomb case in SH approach}
\label{sec:coul-sh}


With good accuracy, one can calculate the amplitude only for the $Y$ coordinate
and then double the result. This is not difficult to prove, but tedious, so we
do not provide a proof here. The amplitude $A$ for the $Y$ coordinate [see Eq.\
(\ref{eq:r3-xy})] from the transition operator Eq.\ (\ref{eq:trans-oper}) with
extracted by Eqs.\ (\ref{eq:coul-ass}) and (\ref{eq:d-coef}) low-energy
dependence is written in terms of the overlap integral $I^{(cc)}$ as
\begin{eqnarray}
A(E_x,E_y) = M_{\text{a}}  D_{l_x}(\eta_x,k_x) \, D_{l_y}(\eta_y,k_y) \,
I^{(cc)}_{l_xl_y}(\varepsilon,E_T) \,,\nonumber \\
I^{(cc)}_{l_xl_y}(\varepsilon,E_T) = \int dX dY
\frac{F_{l_x}(k_xX)}{D_{l_x}(\eta_x,k_x)}
\frac{F_{l_y}(k_yY)}{D_{l_y}(\eta_y,k_y)} Y \psi_{\text{gs}}(X,Y)  , \,
\label{eq:ikfcc}   \\
\eta_x = Z_2Z_3 e^2 M_x/k_x \,, \quad \eta_y = (Z_2+Z_3)Z_1 e^2 M_y/k_y \,.
\nonumber
\end{eqnarray}
The asymptotic form of this overlap, independent of energy, is
\begin{eqnarray}
\tilde{I}^{(cc)}_{l_xl_y} & = & \int dX dY I_{2l_x+1}(2 \sqrt{2 \beta_x X})
I_{2l_y+1}(2 \sqrt{2 \beta_y Y}) \nonumber \\
& \times & \sqrt{XY^3} \, \psi_{\text{gs}}(X,Y) \,.
\label{eq:ikfcc-asym}
\end{eqnarray}
The WF $\psi_{\text{gs}}$ and the integrand of Eq.\ (\ref{eq:ikfcc}) on the
$\{X,Y\}$ plane are shown in Figs.\ \ref{fig:wfs-xy} (a) and (b). Their
comparison illustrates the extreme peripheral character of the low-energy E1
transition: the WF maximum is at a distance of $\sim 3$ fm, while sizable
contributions to the transition ME can be found up to $\sim 60$ fm.

\begin{figure}
\begin{center}
\includegraphics[width=0.45\textwidth]{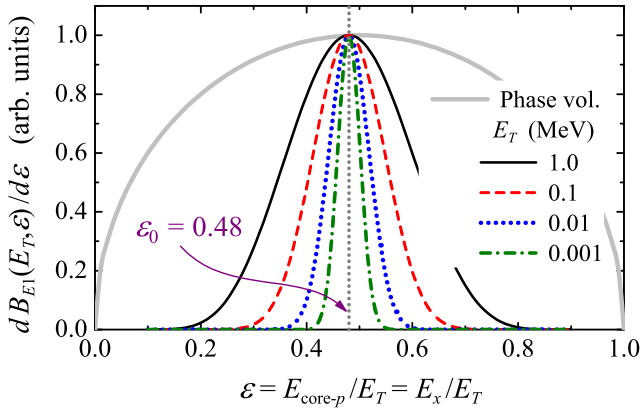}
\end{center}
\caption{Energy distribution between core and one of the protons for
different decay energies $E_T$, governed by the $\exp[-2 \pi(\eta_x +
\eta_y)]$ term in Eq.\ (\ref{eq:ieps-1}). All distributions are normalized to
unity value at peak.}
\label{fig:eps-dis}
\end{figure}

The E1 SF with antisymmetry between nucleons taken into account is
\begin{eqnarray}
\frac{dB_{E1}}{dE_T}  =  G_{fi}\, M^2_{\text{a}} \, 4M_xM_y
\frac{4(Z_3-A_3)^2 e^2 (A_2+A_3)^2}{(A_1+A_2+A_3)^2A^2_3} \,
I_{\varepsilon}(E_T) \,,
\label{eq:dbde2-cou-sh} \\
I_{\varepsilon}(E_T)  = E_T \int_0^1 d \varepsilon I^{(cc)2}_{10}(\varepsilon,
E_T) \exp[-2 \pi (\eta_{x}+\eta_{y})] \,.\quad
\label{eq:ieps-1}
\end{eqnarray}
The Coulomb exponent in $I_{\varepsilon}$ has a very sharp energy dependence,
see Fig.\ \ref{fig:eps-dis}. The energy dependence of $I^{(cc)}$ is shown in
Fig.\ \ref{fig:wfs-xy} (c): it is quite flat for $\varepsilon \approx
\varepsilon_0$. Thus, $I^{(cc)}$ can be evaluated at the
peak $\varepsilon=\varepsilon_0$ and the $\varepsilon$ integration can be
performed by the saddle point method:
\begin{eqnarray}
&  & I_{\varepsilon}(E_T)  =  I^{(cc)2}_{10}(\varepsilon_0, E_T) \,E_T \,
\frac{\exp(-2 \pi \eta_{\text{sh}})}{\sqrt{3 R_{\varepsilon} \eta_{\text{sh}} }}
 \,,
\label{eq:ieps-2} \\
\eta_{\text{sh}} & = & Z_{\text{sh}} e^2 M / \varkappa  \,, \quad Z_{\text{sh}}
= (b_x + b_y)^{3/2} \,,
\label{eq:zeff-sh} \\
\varepsilon_0 & = & b_x/(b_x + b_y)  \,, \quad R_{\varepsilon} = (b_x + b_y)^2/
(4b_xb_y)  \,, \nonumber \\
b_x & = & [Z_3^2Z_2^2M_x/M]^{1/3} \,, \quad b_y = [(Z_3+Z_2)^2Z_1^2M_y/M]^{1/3}
\,. \nonumber
\end{eqnarray}
For the $^{17}$Ne $\rightarrow ^{15}$O+$p$+$p$ transition
\begin{equation}
\varepsilon_0 =  0.48  \,, \quad Z_{\text{sh}}  = 23.282  \,,
\label{eq:zeff-sh-num}
\end{equation}
The accuracy of the saddle point integration is $\sim 2\% $ and $ \sim 6\% $ at
0.1 and 1 MeV, respectively.

It can be found in Fig.\ \ref{fig:overlap-int} that the analytical energy
dependence of Eq.\ (\ref{eq:i2pw-k0}) obtained for the system without Coulomb
interaction is still a good approximation in the considered Coulomb case,
\begin{equation}
I^{(cc)}_{10}(\varepsilon_0,E_T) = \tilde{I}^{(cc)}_{10}
\frac{\sqrt{1+E_T/E_{\text{cy}}}}{(1+E_T/E_{\text{b}})^2}\, , \qquad
E_{\text{cy}}=\frac{2 M_y \beta_y}{1-\varepsilon_0} \,,\;
\label{eq:icc-ener-dep}
\end{equation}
Eq.\ (\ref{eq:icc-ener-dep}) contains additional Coulomb correction for $l_y=1$
motion in $Y$ coordinate (with $E_{\text{cy}}=3.67$ MeV) and we use it later for
astrophysical rate derivation.

The results of the SF calculation in the SH approximation are shown in Fig.\
\ref{fig:e1-sf-17ne} by the red dashed curve. Now there is no significant
disagreement for $E_T \rightarrow 0$ of the SH model results with calculation
results from Refs.\ \cite{Grigorenko:2006} and \cite{Parfenova:2018}. In the
next Sections \ref{sec:correl-2p} and \ref{sec:two-body-res} we answer the
following questions: (i) what is the reason for the difference between HH and SH
ANC3 methods and (ii) can we get a better ``fit'' of the complicated three-body
results in the SH approximation?

\subsection{Correlated $2p$ emission/capture}
\label{sec:correl-2p}


In the HH Eq.\ (\ref{eq:dbde1-cou-hh}) and SH Eq.\ (\ref{eq:dbde2-cou-sh})
approximations we get SF expressions with the low-energy asymptotics
\begin{equation}
\frac{dB_{E1}}{dE_T} \propto \exp(-2 \pi \eta_{\text{hh}}) \,,\quad
\frac{dB_{E1}}{dE_T} \propto E_T^{5/4}  \exp(-2 \pi \eta_{\text{sh}})   \,,
\label{eq:dbde3-qual}
\end{equation}
which are qualitatively different. There are two main points. (i) Effective
charge, entering the Coulomb exponent is significantly lower in SH case, see
Eq.\ (\ref{eq:zeff-sh-num}) compared to Eq.\ (\ref{eq:zhh-k0-k1}): 27.41 vs.\
23.282. (ii) There is an additional power dependence on energy $E^{5/4}_T$
which, evidently, cannot be compensated, for example, by some modification of
the effective charge. What is the source of qualitative difference between Eqs.\
(\ref{eq:dbde1-cou-hh}) and (\ref{eq:dbde2-cou-sh})?

The answer is actually provided in Fig.\ \ref{fig:eps-dis}: in the SH approach
the emission of two protons is highly correlated process, which produces the
narrow bell-shaped $\varepsilon$ distributions. In the approximation $K_f=1$
used in Eq.\ (\ref{eq:dbde1-cou-hh}) the momentum distribution is described by
the phase space (thick gray curve in Fig.\ \ref{fig:eps-dis}). In the correlated
calculation Eq.\ (\ref{eq:dbde2-cou-sh}) this distribution is drastically
modified by the three-body Coulombic effect. The momentum distribution which
``shrinks'' to the proximity of the $\varepsilon_0$ value allows an easier
penetration, which is reflected also in the smaller effective charge
(\ref{eq:zeff-sh-num}) in the Coulomb exponent in Eq.\ (\ref{eq:dbde3-qual}).

So, what is wrong with Eq.\ (\ref{eq:dbde1-cou-hh})? Formally the transition by
the dipole operator from $K_i=0$ g.s.\ occurs only to $K_f=1$ continuum, as
we assumed. This means that the substitution of Eq.\ (\ref{eq:coul-ass-0}) is
incorrect. This substitution is based on the assumption that for $E_T
\rightarrow 0$ only the component with minimal possible $K_f=1$ and, hence, the
smallest centrifugal barrier, contributes to the penetrability. Now it is clear
that for the three-body continuum Coulomb problem this ``evident'' argument is
incorrect. Within the complete HH couple-channel problem the $K_f=1$
channel should be affected by an infinite sum of the other channels in such a
way that their cumulative effect does not vanish even in the limit $E_T
\rightarrow 0$.

Analogous energy correlation effect is well known for the $2p$ radioactivity
process. It was predicted by Goldansky in his pioneering work on $2p$
radioactivity \cite{Goldansky:1960}: in the Coulomb-correlated emission of two
protons the energies of the protons tend to be equal in the limit of
infinitely strong Coulomb interaction in the core+$p$ channel. This effect for
two-proton radioactivity and resonant ``true'' two-proton emission is now well
studied experimentally and understood in details in theoretical calculations
\cite{Pfutzner:2012,Brown:2015}. It is proved that the approximations like Eq.\
(\ref{eq:dbde2-cou-sh}) represent well the underlying physics of the phenomenon.


\subsection{Effect of a resonant state in a two-body subsystem}
\label{sec:two-body-res}


It was shown in calculations of \cite{Grigorenko:2006,Parfenova:2018} (see
Figs.\ 3 and 4-5 in these works) that the resonant state in the core-$p$
subsystem with ``natural parity'' quantum numbers significantly affects
\emph{both} the profile of the E1 strength function in a wide range of
energies and the asymptotic behavior at low $E_T$ values. To evaluate the
influence of resonance on the asymptotics analytically, let us consider the
two-body \emph{resonant} scattering WF in the \emph{quasistationary
approximation}:
\begin{equation}
f_l(kr) = e^{i \delta_l} F_l(kr)  \cos(\delta_l) +
\frac{\sqrt{v\,\Gamma(E)}/2}{E_r-E-i\Gamma(E)/2} \tilde{\psi}_{l}(r)\,. \;
\label{eq:qs-wf}
\end{equation}
This expression can be easily connected with the asymptotics Eq.\
(\ref{eq:con-wf}) by using the resonant R-matrix formulas:
\begin{equation}
\tan (\delta_l) = \frac{\Gamma(E)}{E_r-E} \; \rightarrow \;
e^{i \delta_l} \sin (\delta_l) = \frac{\Gamma(E)/2}{E_r-E+i \Gamma(E)/2} \,.
\label{eq:delta-res}
\end{equation}
The $\tilde{\psi}_{l}(r)$ is so-called \emph{quasistationary} WF, defined at
resonant energy $E_r$ by the irregular Coulomb WF boundary
condition and normalized to unity in the ``internal region'' $r < r_c$:
\begin{equation}
\tilde{\psi}_{l}(r_c) \stackrel{E=E_r}{ \propto} G_l(k_r r_c)\,, \quad
\int_0^{r_c} dr |\tilde{\psi}_{l}(r)|^2 = 1 \,.
\label{eq:quasistat-con}
\end{equation}

The low-energy behavior of the overlap integrals Eq.\ (\ref{eq:ikfcc}) with the
resonant continuum WF (\ref{eq:qs-wf}) in $X$ coordinate is then
\begin{eqnarray}
I^{(cc)\prime }_{l_xl_y}(\varepsilon_0, E_T) = I^{(cc)}_{l_xl_y}(\varepsilon_0,
E_T) + \frac{B(\varepsilon_0,E_T)}{1-\varepsilon_0 E_T/E_r} \,, \quad
\label{eq:icc-corr} \\
B(\varepsilon_0,E_T)  =  B_c   \int dX dY
\tilde{\psi}_{l_x}(X)\frac{F_{l_y}(\eta_y, k_y Y)} {D_{l_y}(\eta_y,k_y)} Y
\psi_{\text{gs}}(X,Y) \,,
\label{eq:b-int} \\
 B_c = \theta_x \, [4M_x r_c K_{2l_x+1}(2\sqrt{2 \beta_x r_c}) E_r]^{-1}  \,.
 \quad \nonumber
\end{eqnarray}
Here we use the R-matrix width definition
\begin{equation}
\Gamma(E) = \frac{\theta^2}{Mr^2_c} \, P_l(E,r_c)\,, \;
P_l(E,r_c) = \frac{kr_c}{F^2_l(kr_c)+G^2_l(kr_c)} \,,
\label{eq:r-matr}
\end{equation}
which is simplified in the low-energy region using Eq.\  (\ref{eq:coul-g-ass}).

The integrand of Eq.\ (\ref{eq:b-int}) is shown in Fig.\ \ref{fig:wfs-xy} (d)
and it has quite peripheral character compared to the g.s. WF Fig.\
\ref{fig:wfs-xy} (a). The ``resonance correction function'' $B$ is shown in
Fig.\ \ref{fig:overlap-int} demonstrating very weak dependence on energy. It is
evaluated with function $\tilde{\psi}_{l_x}(X)$ approximated by Hulten Ansatz
with rms $X$ value 3.5 fm. Parameters $\theta_x=1$ and $r_c=3.7$ fm allows to
reproduce correctly the experimental width $\Gamma = 25(5)$ keV of the $^{16}$F
$0^-$ ground state at $E_r=535$ keV. So, in the whole energy range of interest
we can approximate $B$ as
\begin{equation}
\tilde{B} = B(\varepsilon_0,E_T \rightarrow 0) \,.
\label{eq:tilde-b}
\end{equation}
The blue dotted curve in Fig.\ \ref{fig:e1-sf-17ne} shows nice agreement of the
``resonance corrected'' E1 SF with complete three-body calculations up to $\sim
600$ keV. At this energy the two-body resonance well enters the ``energy
window'' for three-body capture $E_r<E_T$ and turnover to sequential capture
mechanism is taking place.


\subsection{Astrophysical rate calculations}


The \emph{nonresonant} radiative capture rate is expressed via the SF of E1 
dissociation Eq.\ (\ref{eq:dbe1-def}) as
\begin{eqnarray}
\left\langle \sigma _{2p,\gamma }v\right\rangle &= & \left( \frac{\textstyle
\sum_n A_n}{A_1 A_2 A_3}\right)
^{3/2}\left( \frac{2\pi }{mkT}\right) ^{3}\,\frac{2J_f+1}{2(2J_i+1)} \,
I_E(T)\,, \; \nonumber
\\
I_E(T) & = & \int dE \, \frac{16\pi}{9}\, E_{\gamma }^{3}\;
\frac{dB_{E1}(E)}{dE} \exp \left[ -\frac{E}{kT}\right] \,,
\label{eq:nonres-rate}
\end{eqnarray}
where $J_i$ and $J_f$  are spins of the $^{15}$O and $^{17}$Ne g.s.,\
respectively \cite{Parfenova:2018}.

\begin{figure}
\begin{center}
\includegraphics[width=0.49\textwidth]{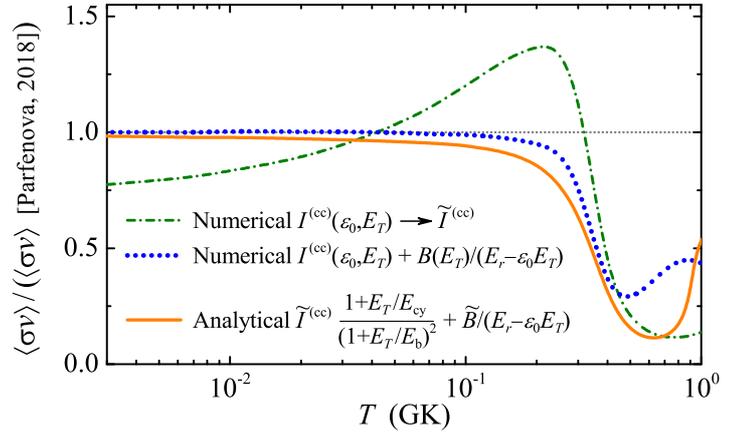}
\end{center}
\caption{Ratio of the rates for astrophysical nonresonant three-body capture
reaction $^{15}$O+$p$+$p \rightarrow ^{17}$Ne+$\gamma$ obtained in this work and
in the paper \cite{Parfenova:2018} (see the thin magenta solid line  SF in Fig.\
\ref{fig:e1-sf-17ne}). Blue dotted and solid orange curve show the results of
numerical rate calculation by Eq.\ (\ref{eq:nonres-rate}) and analytical result
by Eq.\ (\ref{eq:saddle}) for the resonance-corrected
$I^{(cc)\prime}(\varepsilon,E_T)$ from Eq.\ (\ref{eq:icc-corr}). Green
dash-dotted curve shows the rate for simple energy-independent SF
$I^{(cc)}(\varepsilon,E_T) \rightarrow \tilde{I}^{(cc)}$
from Eq.\ (\ref{eq:ikfcc-asym}).}
\label{fig:rate-all}
\end{figure}

The energy dependence of Eq.\ (\ref{eq:dbde2-cou-sh}) is too complex to allow a
direct analytical calculation of the astrophysical capture rate. However, using
Eqs.\ (\ref{eq:ieps-2}), (\ref{eq:icc-ener-dep}), and (\ref{eq:tilde-b}), the
 main analytical terms can be obtained by the saddle point calculation near the
 Gamow peak energy $E_G$:
\begin{eqnarray}
I_E(T) \propto \int d E_T  (E_{\text{b}} + E_T)^3 I_{\varepsilon}(E_T) \exp
\left[- \frac{E_T}{kT} \right]   = \frac{2 \pi E_{\text{b}}^3
E_{\text{G}}^{5/2}}{3 \gamma \sqrt{R_{\varepsilon}}}
\nonumber \\
\times  \frac{1 + E_{\text{G}}/E_{\text{cy}}} {1 + E_{\text{G}}/E_{\text{b}}}
\left( \tilde{I}^{(cc)}_{10} + \frac{(1+E_{\text{G}}/E_{\text{b}})^2}
{1-\varepsilon_0 E_{\text{G}}/E_r} \tilde{B} \right)^2 \exp \left[- \frac{3
\gamma^{2/3}}{(kT)^{1/3}} \right], \, \nonumber \\
E_{\text{G}} = (\gamma k T )^{2/3} \,, \quad \gamma = \pi Z_{\text{sh}} e^2
\sqrt{M/2} \,, \quad \pi \eta_{\text{sh}} = \gamma /\sqrt{E_T} \,.   \quad
\label{eq:saddle}
\end{eqnarray}
The Gamow peak energy can be found as $\{0.05,0.21,1\}$ MeV at $\{0.01,0.1,1\}$
GK. Comparison of the rates calculated in a model of the paper Ref.\
\cite{Parfenova:2018} and in this work is given in Fig.\ \ref{fig:rate-all}.
It can be seen that even the very crude energy-independent approximation Eq.\
(\ref{eq:ikfcc-asym}) $I^{(cc)}(\varepsilon,E_T) \rightarrow \tilde{I}^{(cc)}$
for $T<0.5$ GK works well within a factor of 2. The energy-dependent
calculations Eq.\ (\ref{eq:icc-corr}) are nearly perfect for $T<0.2$ GK and the
analytical expression Eq.\ (\ref{eq:saddle}) for the rate is a very good
approximation to numerically computed rate for $T<0.4$ GK.

The low-temperature dependence of the \emph{nonresonant} $2p$ capture rate is
\[
\langle \sigma _{2p,\gamma}\, v \rangle \propto \mathcal{C}^2_{3} \, T^{5/3} \,
\exp[- (T_{\text{eff}}/T)^{1/3} ] \,, \quad kT_{\text{eff}} \approx 193 \;
\text{MeV} \, .
\]
Analogous dependence for the \emph{resonant} rate is (e.g.\ Ref.\
\cite{Grigorenko:2005a})
\[
\langle \sigma _{2p,\gamma}\, v \rangle \propto \Gamma_{2p} \, T^{-3} \,
\exp[- (E_{3r}/kT)] \,, \quad E_{3r} = 0.355 \; \text{MeV} \, ,
\]
where $E_{3r}$ is the lowest state decaying via $2p$ emission with $\Gamma_{2p}$
(for $^{17}$Ne this is the first excited $3/2^-$ state). Here it can be found
that nonresonant capture dominates in the low-temperature limit, see also
discussion in Ref.\ \cite{Grigorenko:2006}.

So, we have obtained a compact analytical expression for the $2p$ capture rate,
which depends only on the global parameters of the system under consideration
($\mathcal{C}_{3}$, $E_{\text{b}}$, $E_r$, $Z_{\text{sh}}$) and two universal
overlaps ($\tilde{I}^{(cc)}$ and $\tilde{B}$) calculated at $E_T \rightarrow 0$.

\section{General note on three-body Coulomb continuum problem}
\label{sec:coul-3b}


The \emph{three-body Coulomb continuum problem} is a famous long-term conundrum 
of theoretical and mathematical physics. Complexity of this problem is defined 
by the possible presence of the Coulomb correlations, bound and resonant states 
in the two-body subsystems. As a result, no compact analytical form of the 
asymptotics is known for the three-body Coulomb continuum problem. There is 
known approximate asymptotic solution of this problem (so-called 
``Redmond-Merkuriev'' asymptotics) \cite{Rosenberg:1973,Merkuriev:1977}, which 
is valid in \emph{four} regions: in \emph{one} region all three particles are 
far from each other $r_{12} \sim r_{23} \sim r_{31}$ and there are \emph{three} 
regions where different pairs of three particles are close $r_{ij} \ll r_{jk} 
\sim r_{ki}$. There were fruitful applications of this asymptotics, e.g.\ to 
atomic problems \cite{Peterkop:1977,Brauner:1989}. There is a wide range of 
works dedicated to improvement of this asymptotics 
\cite{Alt:1993,Mukhamedzhanov:2006b,Kadyrov:2009,Yakovlev:2020}. One of modern 
trends is not to struggle with analytical problems of this asymptotics, but to 
use powerful computing and propagate numerical solutions to distances where 
uncertainties of the asymptotics does not play a practical role. However, we do 
not think that this is a completely satisfactory approach, which should replace 
the analytical developments.

\begin{figure}
\begin{center}
%
\includegraphics[width=0.49\textwidth]{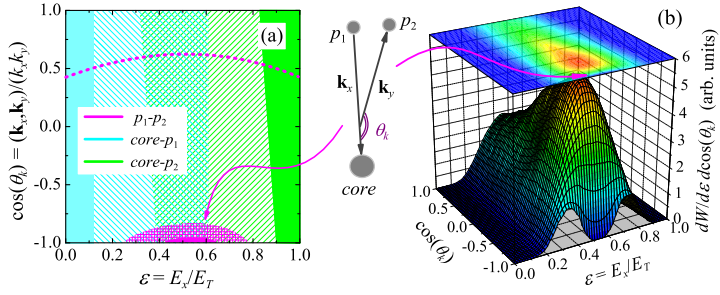}
\end{center}
\caption{(a) Regions in the kinematical plane $\{\varepsilon,\cos(\theta_k)\}$ 
where effects of the two-body Coulomb repulsion are dominating. Different colors 
correspond to different pairs of particles, where $r_{ij} \ll r_{jk} \sim 
r_{ki}$. Solid color corresponds to $E_T \sim 1$ MeV [to be compared  with panel 
(b)], hatched regions to $E_T \sim 1-100$ keV. The dotted magenta curve 
illustrates the region of $p$-$p$ repulsion dominance at some hypothetical 
extremely small energy $E_T \ll 1$ keV. (b) Illustration of the Coulomb 
repulsion dominance regions by example of theoretical momentum distribution for 
decay $^{16}$Ne(g.s.)$\rightarrow^{14}$O+$p$+$p$ with $E_T = 1.466$ MeV, very 
well reproducing the data \cite{Brown:2015}.}
\label{fig:coul3-ill}
\end{figure}

In this work we deal with a limited subset of the three-body Coulomb continuum 
problem: only repulsive Coulomb interactions and no bound states in the 
subsystems. Specific feature of our problem is that the three-body energies are 
extremely small and the solution residue in the kinematical region, where the 
contributions of two two-body Coulomb asymptotics (core-$p_1$ and core-$p_2$ 
channels) overlap, see Fig.\ \ref{fig:coul3-ill} (a). This justifies amplitude 
factorization and consequent analytical calculations of Section 
\ref{sec:coul-sh}. However, reliability of this approximation is based on the 
fact that $p$-$p$ Coulomb interaction is quite small compared to core-$p$ 
interaction. It can be found from Fig.\ \ref{fig:coul3-ill} (b) that a minor 
part of the kinematical space is affected by the $p$-$p$ repulsion for energies 
$1<E_T<1000$ keV important for astrophysics. It should be understood, that, 
rigorously speaking, for some extreme $E_T \rightarrow 0$ the correlations 
induced by $p$-$p$ Coulomb interaction will become important in the whole 
kinematical plane $\{\varepsilon,\cos(\theta_k)\}$ and the low-energy 
asymptotics, which we deduced in this work, will be broken.

\section{Conclusion}
\label{sec:concl}


In this paper, we provide a formalism for a complete analytical description of
low-energy three-body $2p$ nonresonant radiative capture processes. The
developed approach is a generalization of the ANC method, which has proven
itself well for two-body nonresonant radiative captures. The ordinary (two-body)
ANC2 method demonstrates the sensitivity of the low-energy E1 strength function,
important for astrophysics, to only two parameters: the binding energy
$E_{\text{b}}$ and the ANC2 value $\mathcal{C}_2$. For the three-body ANC3
method
one more parameter should be employed: the energy $E_r$ of the lowest to the
threshold ``natural parity'' two-body resonance with appropriate quantum
numbers.

An interesting formal result is related to the problem of the three-body Coulomb
interaction in the continuum. We demonstrate that ANC3 method developed
completely in three-body hyperspherical harmonics representation (named ``HH
approximation'') is not valid in the Coulomb case as it gives incorrect
low-energy asymptotics of SF and hence incorrect low-temperature asymptotics of
the astrophysical rate. The reason for this is the highly correlated nature of
the $2p$ capture. The correct asymptotics can be obtained using the
Coulomb-correlated SH approximation based on a simplified three-body
Hamiltonian. The latter approximation also allows to determine the correction
related to the low-lying two-body resonant state in the core-nucleon channel.

The two-dimensional overlap integrals involved in the ANC3 approximation in the
correlated Coulomb case are rather complicated compared to those in the ANC2
case. However, their calculation is a task that is incomparably simpler than any
complete three-body calculation. The whole formal framework is compact and
elegant and requires only two overlap calculations: $\tilde{I}^{(cc)}$ and
$\tilde{B}$. Thus, we find that ANC3 approximation in a three-body case is a
valuable development providing robust tool for estimates of the three-body
nonresonant capture rates in the low-temperature ($T\lesssim 0.3-1$ GK) domain.

%
\textit{Acknowledgments.}
%
%
--- LVG, YLP, and NBS were supported in part by the Russian Science
Foundation grant No.\ 17-12-01367.



\end{document}